**High-Q Lithium Niobate Microring Resonator with Electro-Optically Reconfigurable Coupling Strength**

*Yuan Ren, Yong Zheng\*, Ruixue Liu, Boyang Nan, Ziyao Zhuang, Rongbo Wu\*, Min Wang, and Ya Cheng\**

Yuan Ren, Yong Zheng, Rongbo Wu, Min Wang, Prof. Ya Cheng
Shanghai Research Center for Quantum Sciences, Shanghai 201315, China
E-mail: yzheng@phy.ecnu.edu.cn, rbwu@phy.ecnu.edu.cn, ycheng@phy.ecnu.edu.cn

Yuan Ren, Ruixue Liu, Boyang Nan, Prof. Ya Cheng
State Key Laboratory of Precision Spectroscopy, East China Normal University, Shanghai 200062, China

Yuan Ren, Yong Zheng, Ruixue Liu, Boyang Nan, Rongbo Wu, Min Wang, Prof. Ya Cheng
The Extreme Optoelectromechanics Laboratory (XXL), School of Physics, East China Normal University, Shanghai 200241, China

Ziyao Zhuang
WLSA Shanghai Academy, Shanghai 200940, China

Prof. Ya Cheng
Hefei National Laboratory, Hefei 230088, China
Collaborative Innovation Center of Extreme Optics, Shanxi University, Taiyuan 030006, China

Funding: Project supported by Shanghai Municipal Science and Technology Major Project (Grant No.2019SHZDZX01). National Natural Science Foundation of China (12192251, 12334014, 62335019, 12134001, 12304418, 12474378). Quantum Science and Technology-National Science and Technology Major Project (2021ZD0301403).

The development of sophisticated integrated photonic circuits demands microresonators that combine exceptional optical confinement with dynamic operational flexibility. Here, we demonstrate a racetrack resonator on the thin-film lithium niobate platform that achieves an electro-optically tunable coupling strength while maintaining a stable, high intrinsic Q factor on the order of $10^6$. By incorporating a Mach-Zehnder interferometer into the coupling region, the device facilitates a continuous and reversible transition across the entire coupling spectrum—from under-coupling and critical coupling to deep over-coupling. To ensure high spectral purity, we employ Euler bends to facilitate an adiabatic transition between the straight and curved waveguide sections. This design effectively suppresses the excitation of higher-order modes, resulting in a clean transmission spectrum characterized by exclusive fundamental mode operation. At the critical coupling point, the resonator exhibits a high extinction ratio exceeding 30 dB. The integration of stable ultra-high Q, single-mode purity, and full-range coupling reconfigurability positions this device as a vital component for adaptive microwave photonics, high-efficiency nonlinear optics, and programmable quantum photonic networks.

# 1.Introduction

Whispering gallery mode resonators and integrated micro-ring resonators have emerged as fundamental building blocks of modern integrated photonics, owing to their exceptional quality(Q) factors and tight modal confinement. By trapping photons within micron-scale volumes and significantly enhancing light-matter interactions, these devices provide a versatile platform for a wide array of applications: ranging from high-speed, low-power comsumption optical communications and photonic computing[1, 2] to nonlinear optical applications such as microcomb generation and frequency conversion,[3-5] as well as serving as efficient single-photon sources and interfaces for quantum information processing.[6, 7] High-Q optical microcavities have been demonstrated on diverse material platforms such as silicon,[8, 9] silicon nitride,[10, 11] and lithium niobate.[12-19] Nevertheless, achieving precise, stable, and arbitrary dynamic control over the coupling state while maintaining an ultra-high Q factor remains a critical bottleneck for the realization of fully reconfigurable photonic systems.

Currently, the mainstream schemes of coupling microring or microdisk resonators using tapered fibers or proximity waveguides are constrained by static microresonator structures defined after the fabrication stage, rendering the coupling ratio unalterable once processing is complete. However, integrated photonic systems in practical operation face diverse and mutually exclusive performance requirements. For instance, electro-optic microring modulators need to be strictly locked in the "critical coupling" state to achieve maximum extinction ratios.[20] Conversely, in soliton microcomb generation, the device typically needs to be precisely positioned in the "over-coupling" state to optimize output power and energy conversion efficiency while simultaneously reducing the pumping threshold.[21-24] This irreversibility of the coupling mechanism severely limits the versatility of high-performance microcavities in complex, adaptive integrated systems.

To overcome the limitations of static coupling, thin-film lithium niobate (TFLN) presents a decisive solution by leveraging its strong intrinsic Pockels effect to facilitate dynamic and precise tuning of the coupling state. The strong Pockels effect in TFLN makes it suitable for realizing high-speed, power-efficient, and low-crosstalk phase shifters which are key components in high-speed optical modulators and advanced photonic networks. [25-35] More importantly, TFLN allows for the monolithic integration of high-speed and power-efficient electro-optic interferometric structures with microrings while maintaining ultra-low optical loss. This unique combination of physical attributes provides the physical possibility to achieve dynamic, precise, and continuous arbitrary transitions from under-coupling and critical coupling to over-coupling without introducing additional absorption loss.

In this work, we demonstrate a high-performance integrated racetrack resonator based on the thin-film lithium niobate (LNOI) platform. The device achieves arbitrary dynamic tuning of the coupling ratio while maintaining a stable high Q factor on the order of $10^6$. By integrating an electro-optically controlled Mach-Zehnder interferometer (MZI) coupler, we are able to precisely drive the system through a continuous transition from the under-coupling to the over-coupling state without sacrificing the intrinsic loss of the resonator. Notably, the device exhibits exceptional mode purity characterized by pure fundamental mode excitation with no higher-order mode interference and delivers a high extinction ratio exceeding 30 dB. The design which converges an ultra-high Q factor, robust fundamental mode excitations, and full-range coupling tunability, holds great promise as a powerful tool for the realization of ideal critical coupling and facilitating the construction of large-scale reconfigurable photonic networks.

## 2. Design Principle

Figure 1(a) illustrates the integrated racetrack resonator, incorporating a Mach-Zehnder interferometer (MZI) in the coupling region to enable arbitrary control of the coupling ratio. To minimize scattering losses and suppress the excitation of higher-order modes, directional couplers (DCs) which exhibit adiabatic-like modal evolution, are employed as beam-splitting elements within the MZI. However, due to inevitable fabrication tolerances, the splitting ratio of the DCs often deviates from the ideal 50:50, which limits the achievable extinction ratio of a conventional single-stage MZI. To circumvent this, we use a cascaded dual-stage Mach-Zehnder interferometer (2-MZI) architecture by incorporating an additional phase shifter (PS) and a directional coupler, thereby enabling "perfect MZI" functionality through interferometric compensation. The structure of this perfect MZI is shown in Figure 1(b). In the following, we analyze the working principle of the architectures by considering their transmission matrices.

The transmission matrix of a single DC with a splitting ratio of $R$ can be expressed as:

$$T_{DC}(R) = \begin{bmatrix} \sqrt{R} & i\sqrt{1-R} \\ i\sqrt{1-R} & \sqrt{R} \end{bmatrix}, \qquad (1)$$

The transmission matrix of a single PS with a relative phase difference $\theta$ can be expressed as:

$$T_{PS}(\theta) = \begin{bmatrix} e^{i\frac{\theta}{2}} & 0 \\ 0 & e^{-i\frac{\theta}{2}} \end{bmatrix}. \qquad (2)$$

Thus, the transmission matrices for a cascaded 2-MZI unit as shown in Figure 1 (b) can be derived as follows:

$$T_{2-MZI} = T_{DC}(R_R) \cdot T_{PS}(\theta_R) \cdot T_{DC}(R_M) \cdot T_{PS}(\theta_L) \cdot T_{DC}(R_L). \qquad (3)$$

Based on the transmission matrix analysis, the 2-MZI architecture facilitates an exceptional extinction ratio even when the directional coupler splitting ratio deviates from the ideal 50:50, thereby enabling arbitrary and precise tuning of the coupling coefficient. In our device, all phase shifters are driven by the electro-optic effect of lithium niobate (LN). The Z-axis of the LN crystal is oriented within the horizontal plane and perpendicular to the waveguide. Under this configuration, when the voltage is applied, the ground-signal-ground electrodes generate electric fields of equal magnitude but opposite directions in the two arms of each MZI, thereby introducing the desired phase difference between the light propagating in the two arms. The simulated optical and electric field distributions are depicted in Figure 1(c), where the electrodes are 500 nm thick, spaced 6.5 μm apart, and fabricated on a 1.5-μm-thick silicon dioxide layer.

To experimentally demonstrate the coupling behavior of the racetrack resonator, our device is fabricated on a X-cut 4-inch LNOI wafer with a 500-nm-thick LN layer bonded to a buried silica layer on the 500-μm-thick silicon support (NANOLN) using PLACE technology. Details of the fabrication process can be found in our previous work.[36] The fabricated programmable racetrack resonator with an MZI phase-shifting arm length of 5 mm is depicted in Figure 2(a). In general, the Q factor of such resonators is mainly determined by the waveguide propagation loss and modal mismatch introduced in the coupling and bending regions. Despite the PLACE technique enables the fabrication of highly uniform, ultra-low-loss photonic devices,[19, 36] maintaining low loss and stable single-mode operation in a microresonator structure still requires careful control of modal compatibility between straight and curved waveguide sections.

Figures 2(b) and 2(c) display the simulated optical field distributions of the fundamental TE mode at a wavelength of 1550 nm in a straight waveguide and in a curved waveguide with a bending radius of 220 μm. Although the two mode profiles appear qualitatively similar, the quantitative overlap analysis reveals that a conventional circular bend introduces a non-adiabatic transition at the straight-to-bend interface, resulting in a mode overlap of 0.966. This incomplete overlap indicates that a fraction of the optical power is transferred to higher-order guided modes or radiative modes, which subsequently contributes to excess propagation loss.

To address this issue, Euler bends are introduced to enforce a continuous variation of curvature along the propagation direction, thereby satisfying the adiabatic condition for modal evolution. As a consequence, the straight-to-bend mode overlap approaches 1.0, effectively suppressing parasitic mode excitation and radiation loss. This adiabatic bend design is essential for minimizing the intrinsic loss of the racetrack resonator, preserving strict single-mode excitation, and ultimately enabling the realization of ultra-high-Q resonances.

## 3. Device characterization

To evaluate the Q factor of the fabricated electrically tunable coupled racetrack resonator, we measured its transmission spectrum using the setup depicted in Figure 3(a). Light from a tunable laser (CTL1550, TOPTICA) was passed through an in-line polarization controller (PSY 201, General Photonics Corp.) to set the TE polarization, and then coupled into the chip via a pair of ultra-high numerical aperture fiber arrays. After voltage-controlled tuning, the output light was collected through another fiber array, then fed into a photodetector (APD-2M-A-100K, Luster), and finally recorded and analyzed with a real-time oscilloscope (MSO64B, Tektronix). The voltage control signals were supplied by a 64-channel NI module (PXIe-6739, National Instruments) with a range of ±10 V, and delivered to the on-chip micro-electrodes through a micro-probe array.

Using this measurement setup, we systematically characterized the spectral responses of the resonator across various electro-optically tuned coupling states. As illustrated in Figure 3(b), by independently controlling the driving voltages applied to the two phase-shifting arms of the 2-MZI, the system achieves a continuous and reversible transition from the under-coupling to the over-coupling state. Here, the voltages in the left column are used to control $\theta_L$ and the right one controls $\theta_R$ as shown in Figure 3(b). Notably, at the critical coupling point, the device exhibits exceptional performance, characterized by a quality factor exceeding $10^6$ and an extinction ratio of over 30 dB.

To further evaluate the tuning robustness of this architecture under non-ideal splitting conditions, we plotted the three-dimensional distribution of the extinction ratio for the bar and cross routes as a function of $\theta_L$ and $\theta_R$ based on Eq. (3). As shown in figure 3(c), the results demonstrate that even with a significantly unbalanced directional coupler splitting ratio of 30:70, high extinction ratios can still be maintained in both routes through the coordinated modulation of the two phase-shifters. This inherent capability is pivotal for achieving arbitrary reconfigurability of the coupling strength and also proves the superior fabrication tolerance of the 2-MZI architecture against process variations.

Figure 4(a) shows a typical transmission spectrum measured using the setup illustrated in Figure 3(a), where the wavelength is scanned along the horizontal axis and the corresponding transmitted signal is recorded vertically. The highly regular spectral response confirms successful suppression of higher-order modes, yielding an FSR of 0.041 nm, which is in good agreement with our Lumerical eigenmode simulation result. One of the fundamental mode at the resonant wavelength of 1550.092 nm was chosen for the measurement of the intrinsic Q factor of $1.35 \times 10^6$ in the critical coupling state as evidenced in Figure 4(b). Benefiting from the smooth sidewalls of waveguides fabricated by the PLACE technique and the specifically optimized Euler bend design, we achieved ultra-high-Q fundamental mode excitation over a broad wavelength range, as shown in Figure 4(c)-(g).

To further verify the stability and tunability of the ultra-high Q performance, six identical structures were fabricated on multiple chips, and their Q factors were systematically analyzed.

As summarized in Figure 4(h), all devices exhibit intrinsic Q factors on the order of $10^6$ under electrical tuning, while maintaining access to the critical coupling condition and fundamental mode excitation. The results confirm the robustness of both the design strategy and the fabrication process.

## 4. Conclusion and Discussion

To conclude, we have demonstrated a high-performance racetrack resonator on thin-film lithium niobate featuring electro-optically tunable coupling. The device achieves stable intrinsic Q factors on the order of $10^6$. By independently controlling the phases in the 2-MZI coupler, we achieve continuous and reversible tuning of the coupling condition from under-coupling to over-coupling and precisely target the critical coupling point with an extinction ratio exceeding 30 dB. Furthermore, optimized adiabatic bends ensure excitation of the fundamental mode with effective suppression of higher-order modes.

The combination of high-Q, dynamically tunable coupling, and pure fundamental mode excitation makes this device a versatile core component for advanced photonic circuits. Specifically, our design presents the potential to function as a reconfigurable high-selectivity filter,[37] a programmable optical buffer for dynamic delay generation,[38] or a nonlinear optical neuron with a tunable activation threshold for photonic neural networks.[39-41] This work is expected to provide a robust and fully controllable building block, potentially paving the way for more intelligent and adaptable integrated photonic systems.

**Acknowledgements**

The work was supported by the Shanghai Municipal Science and Technology Major Project (Grant No.2019SHZDZX01). National Natural Science Foundation of China (12192251, 12334014, 62335019, 12134001, 12304418, 12474378). Quantum Science and Technology-National Science and Technology Major Project (2021ZD0301403).

**Data Availability Statement**

The data that support the findings of this study are available on request from the corresponding author. The data are not publicly available due to privacy or ethical restrictions.

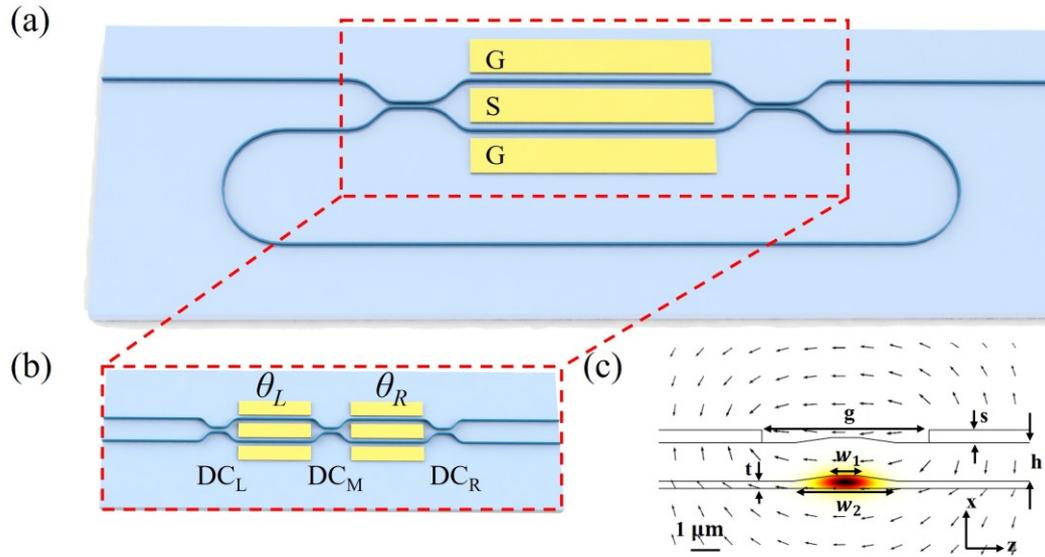

**Figure 1.** (a) Schematic of the racetrack resonator integrated with Mach–Zehnder interferometers. (b) The architectures of 2-cascaded MZIs for performing "perfect MZI" with ultimate high extinction ratios. (c) The simulated optical mode profile and electrical field shown by black arrows (g = 6.5 µm, $w_1$ = 1.0 µm, $w_2$ = 4.8 µm, s = 500 nm, t = 290 nm, h = 1.5 µm).

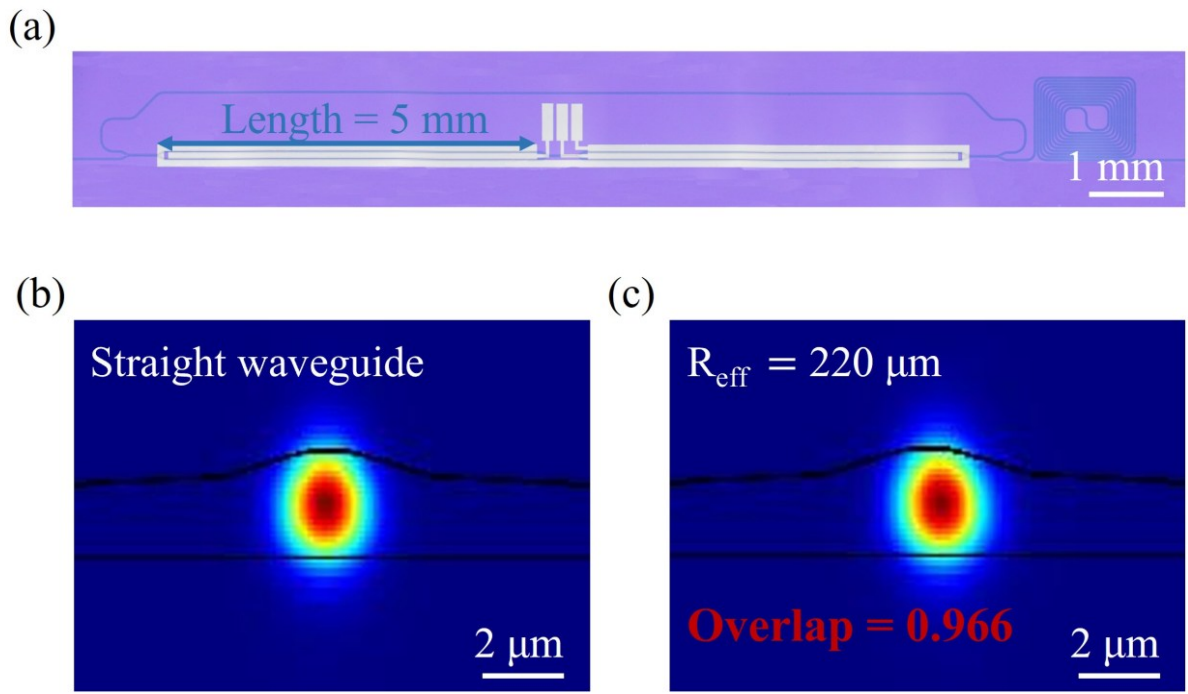

**Figure 2.** (a) Optical microscope image showing a racetrack resonator with an MZI phase-shifter arm length of 5 millimeters. (b) Numerically simulated optical field profile of the fundamental TE mode at a wavelength of 1550 nm in the straight waveguide. (c) Numerically simulated optical field profile of the fundamental TE mode at a wavelength of 1550 nm in the curved waveguide with a bend radius of 220 μm.

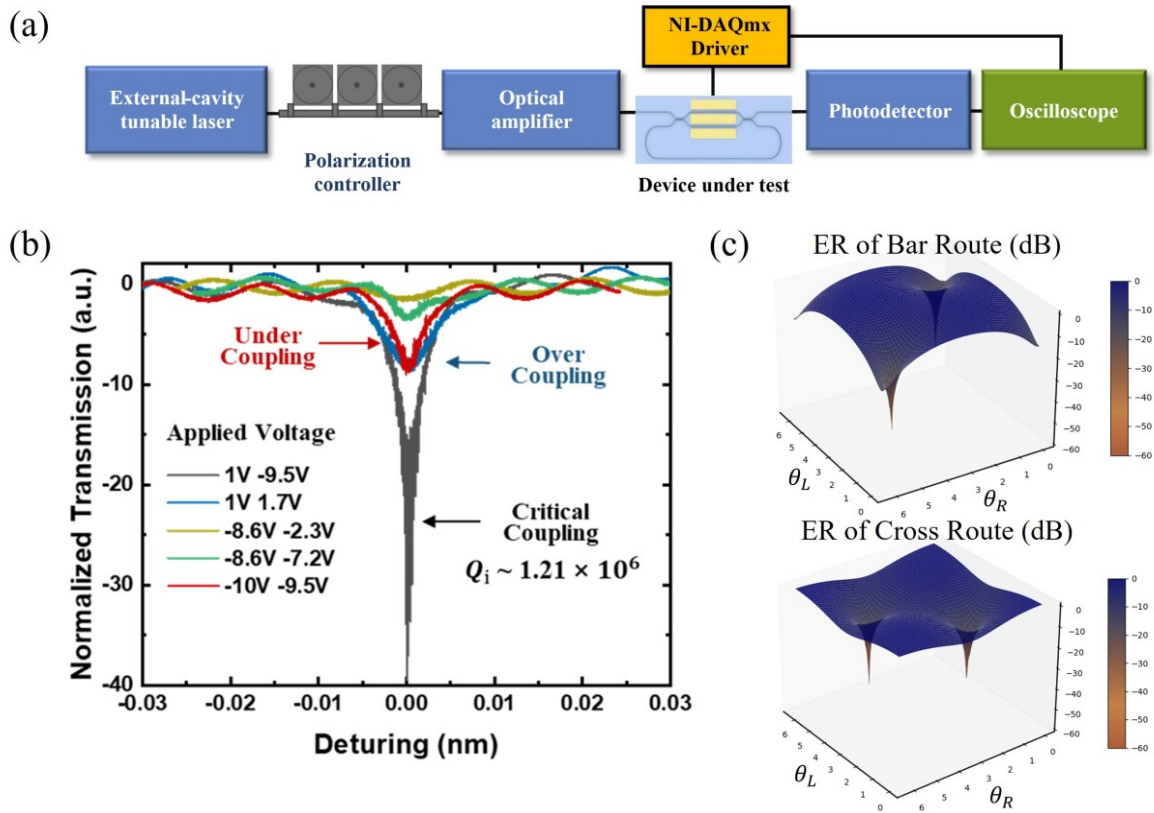

**Figure 3.** (a) Schematic of the experimental setup for measuring the Q factor of the racetrack resonator with electrically tunable coupling states. (b) Experimental transmission spectra of racetrack resonator by voltage tuning. The voltages in the left column are used to control $\theta_L$, and those in the right column are used to control $\theta_R$. Red curve: Under-coupling. Blue curve: Over-coupling. Black curve: Critical coupling. (c) The 3D plots of the ERs of the bar/cross route as functions of $\theta_L$ and $\theta_R$, with all split ratios of DCs set as 30:70.

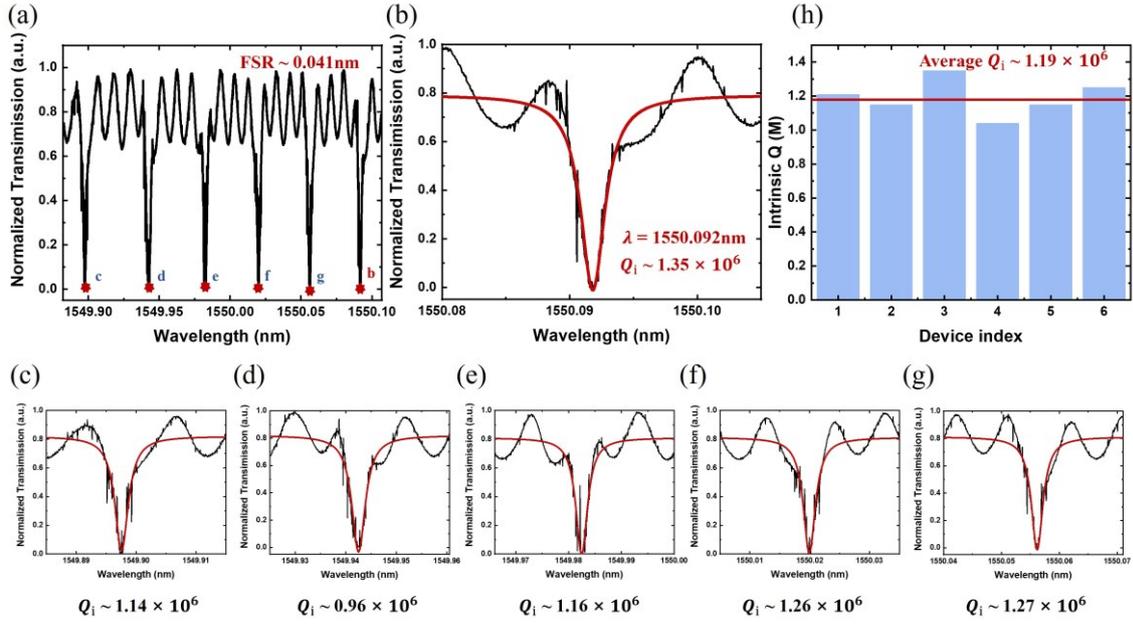

**Figure 4.** (a) Selected resonator spectrum spanning from wavelengths 1549.88 nm to 1550.12 nm. Transmission spectrum without the excitation of higher-order modes. (b) The highest-Q resonance features an intrinsic Q factor of $1.35 \times 10^6$ at the wavelength of 1550.092 nm. Red curve: The Lorentz fitting. (c)–(g) Resonances at wavelengths 1549.899 nm, 1549.941 nm, 1549.982 nm, 1550.020 nm, and 1550.058 nm, all of them belong to the high-Q fundamental mode. (h) Intrinsic Q factors of different groups of racetrack resonators on the different chip.